\documentclass[prd,12pt]{revtex4}

\usepackage{amsmath,amssymb} 
\usepackage[dvips]{graphicx} 
\usepackage[cp1251]{inputenc}
\input{epsf.tex}

\newcommand\nn{\nonumber}
\newcommand\ba{\begin{eqnarray}}
\newcommand\ea{\end{eqnarray}}
\newcommand{\br}[1]{\left( #1 \right)}
\newcommand{\brs}[1]{\left[ #1 \right]}

\newcommand{\brm}[1]{\left| #1 \right|}

\newcommand{\nb}{~\mbox{nb}}

\begin{document}
\date{}

\title{Charge-odd correlation of lepton and pion pair production in
electron-proton scattering}

\author{A.I.~Ahmadov}
\email{ahmadov@theor.jinr.ru}
\affiliation{Institute of Physics, Azerbaijan National Academy of Science, Baku, Azerbaijan}
\affiliation{JINR-BLTP, 141980 Dubna, Moscow region, Russian Federation}

\author{Yu.M.~Bystritskiy}
\email{bystr@theor.jinr.ru}
\affiliation{JINR-BLTP, 141980 Dubna, Moscow region, Russian Federation}

\author{E.A.~Kuraev}
\email{kuraev@theor.jinr.ru}
\affiliation{JINR-BLTP, 141980 Dubna, Moscow region, Russian Federation}

\author{A.~N.~Ilyichev}
\email{ily@hep.by}
\affiliation{National Scientific 
and Educational Centre of Particle and High Energy Physics of the Belarusian State University,
220040 Minsk, Belarus}

\begin{abstract}
Charge-odd correlation of the charged pair components produced at
electron-proton scattering can measure three current correlation
averaged by proton state. In general these type correlation can be
described by 14 structure functions. We restrict here by consideration
of inclusive distributions of a pair components,
which is the light-cone projection of the relevant hadronic tensor.
Besides we consider the point-like approximation for proton and pion.
Numerical estimations show that charge-odd effects can be measured in exclusive
$ep \to 2\pi X$ experiments.
\end{abstract}
\maketitle
%
\section{Introduction}
%
The real photon production in lepton nucleon
scattering allow to extract Deep virtual Compton scattering (DVCS)
amplitude describing the real hard photon emission by proton block.
In such kind of experiments some additional information compared with one obtained in
Deep inelastic scattering (DIS) experiments can be extracted. Among them the important
General Parton Distribution (GPD), \cite{Diehl:2003ny} which generalize the parton (gluon) distribution
inside proton which is extracted from DIS experiments.
\begin{figure}
\unitlength 0.5mm
\begin{tabular}{ccc}
\begin{picture}(80,80)
\put(-10,0){
\epsfxsize=45mm
\epsfysize=45mm
\epsfbox{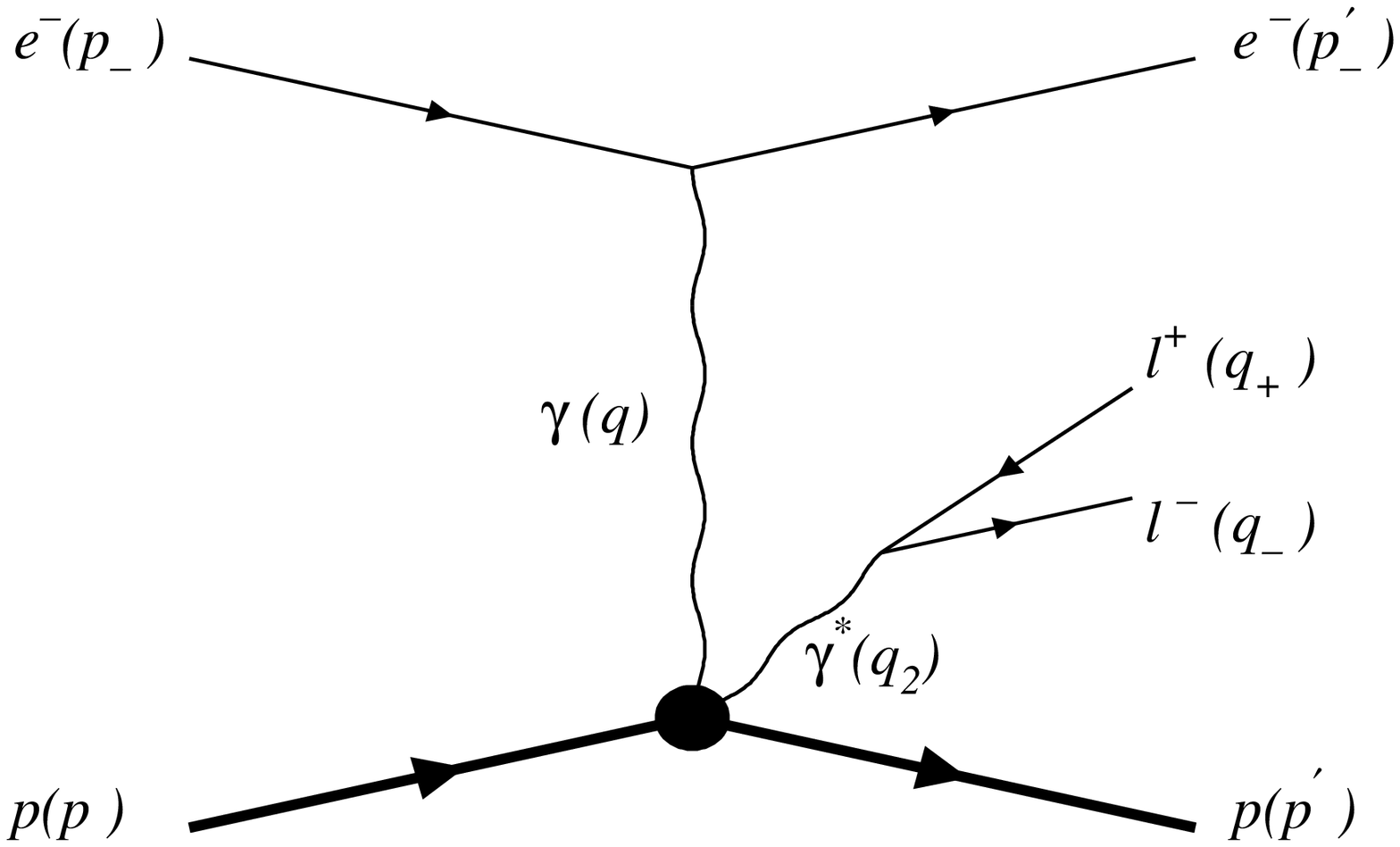}
\put(-50,10){\mbox{a)}}
}
\end{picture}
&
\begin{picture}(80,80)
\put(-5,0){
\epsfxsize=45mm
\epsfysize=45mm
\epsfbox{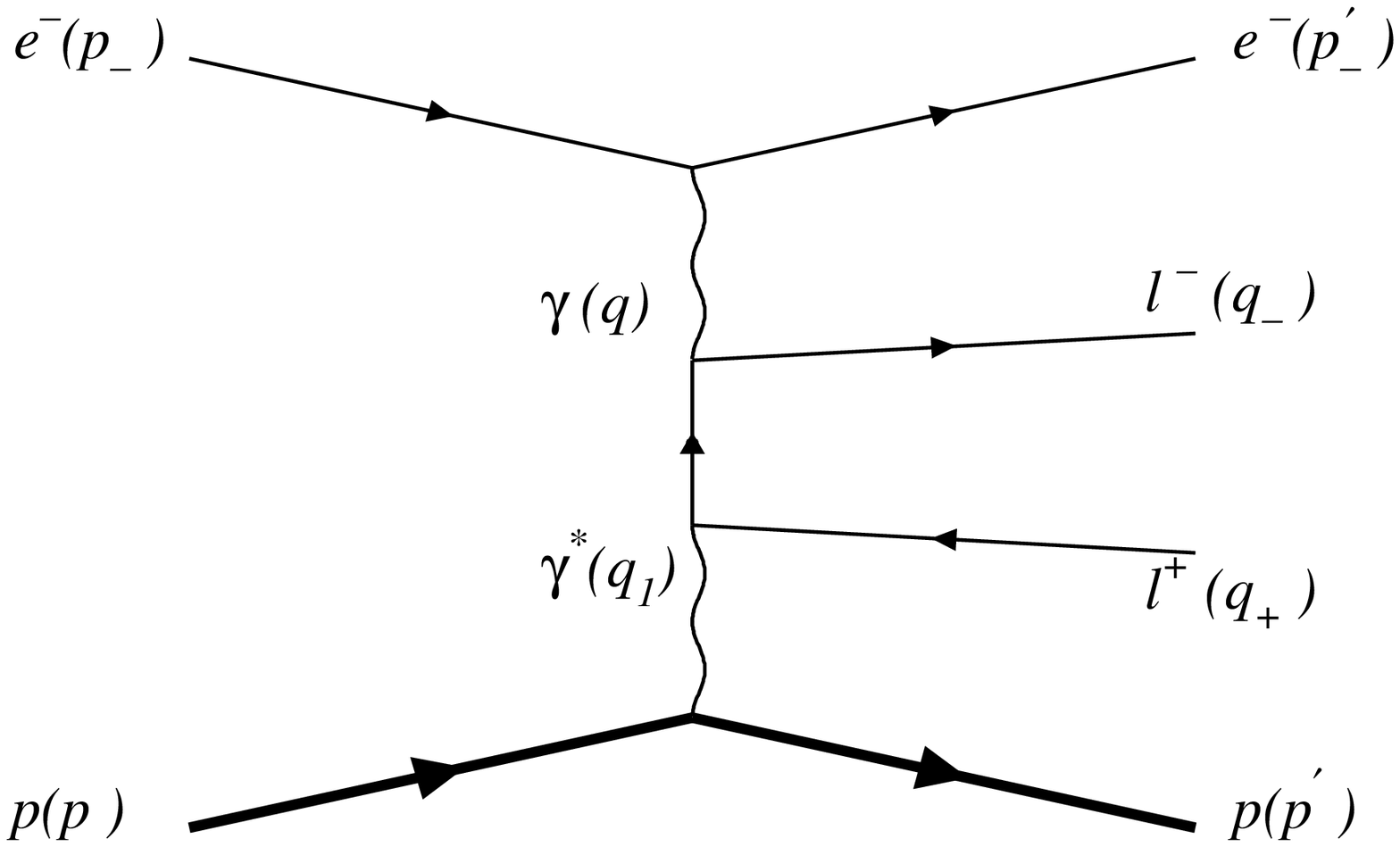}
\put(-50,10){\mbox{b)}}
}
\end{picture}
&
\begin{picture}(80,80)
\put(0,0){
\epsfxsize=45mm
\epsfysize=45mm
\epsfbox{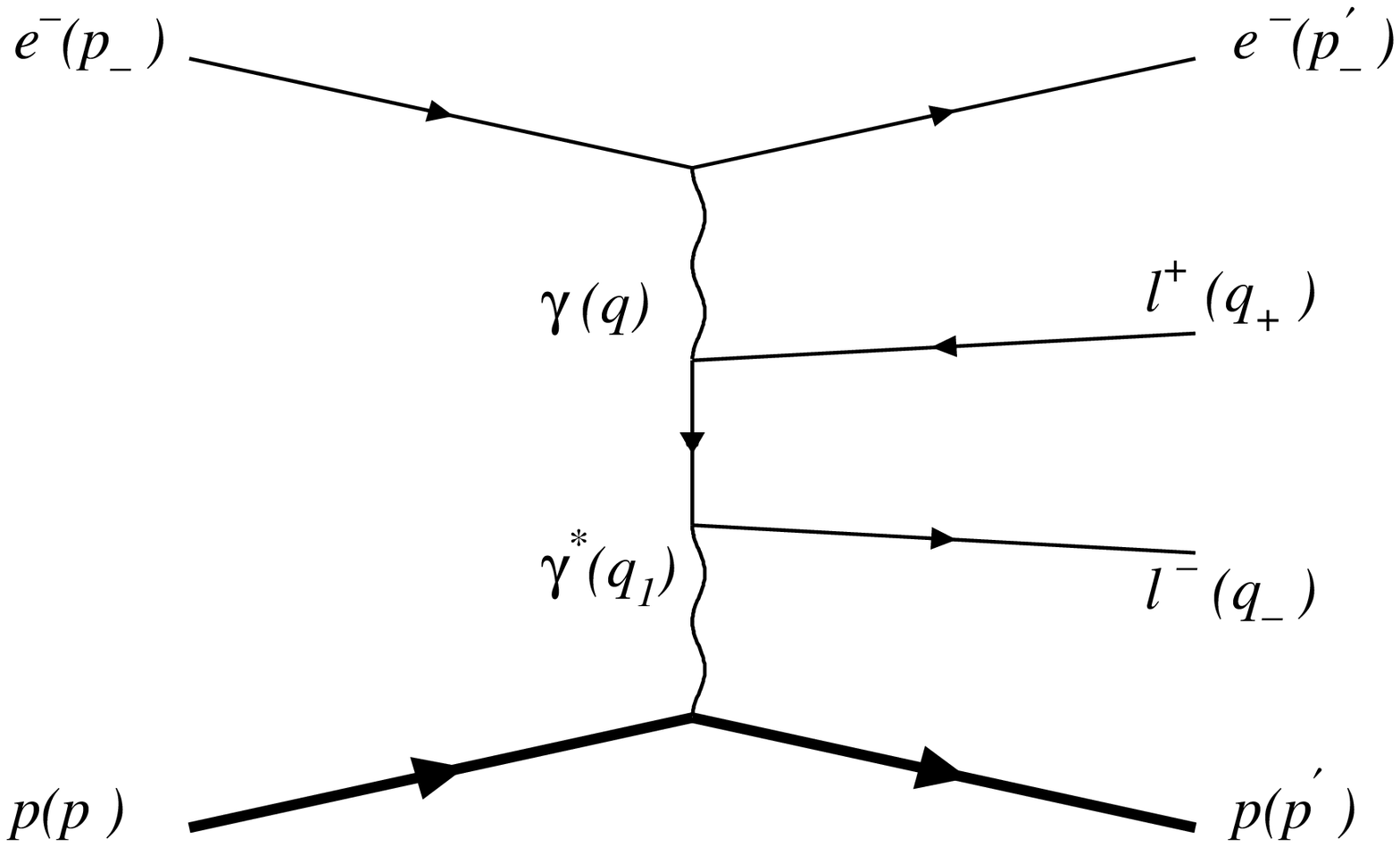}
\put(-50,10){\mbox{c)}}
}
\end{picture}
\\
\begin{picture}(80,80)
\put(-10,0){
\epsfxsize=45mm
\epsfysize=45mm
\epsfbox{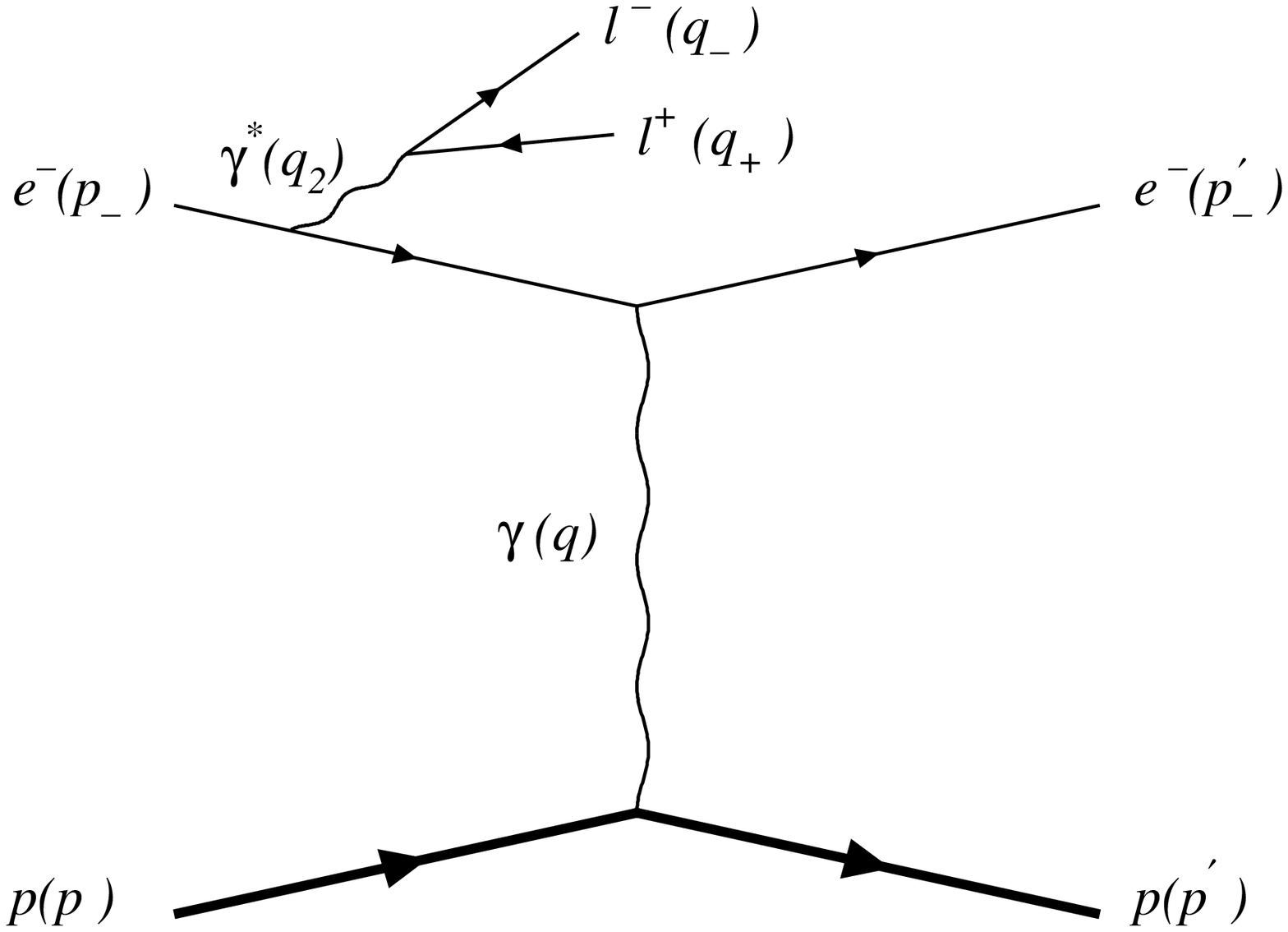}
\put(-50,10){\mbox{d)}}
}
\end{picture}
&
\begin{picture}(80,80)
\put(-5,0){
\epsfxsize=45mm
\epsfysize=45mm
\epsfbox{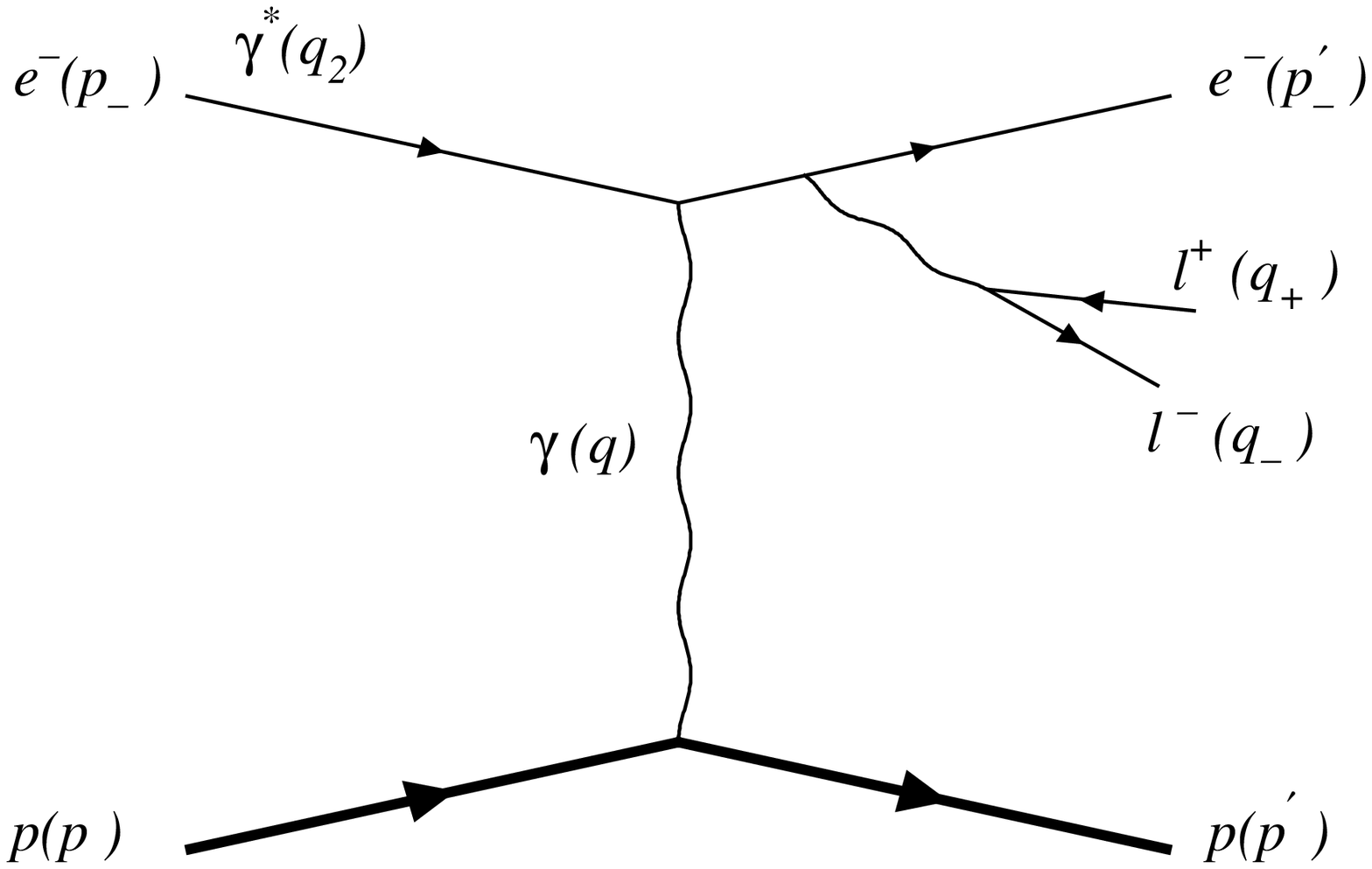}
\put(-50,10){\mbox{e)}}
}
\end{picture}
\end{tabular}
\caption{
Additional lepton pair production in electron-proton scattering:
a) two photon and b)-e) bremsstrahlung mechanisms.
}
\label{fig:pll}
\end{figure}

\begin{figure}
\unitlength 0.5mm
\begin{tabular}{ccc}
\begin{picture}(80,80)
\put(-10,0){
\epsfxsize=45mm
\epsfysize=45mm
\epsfbox{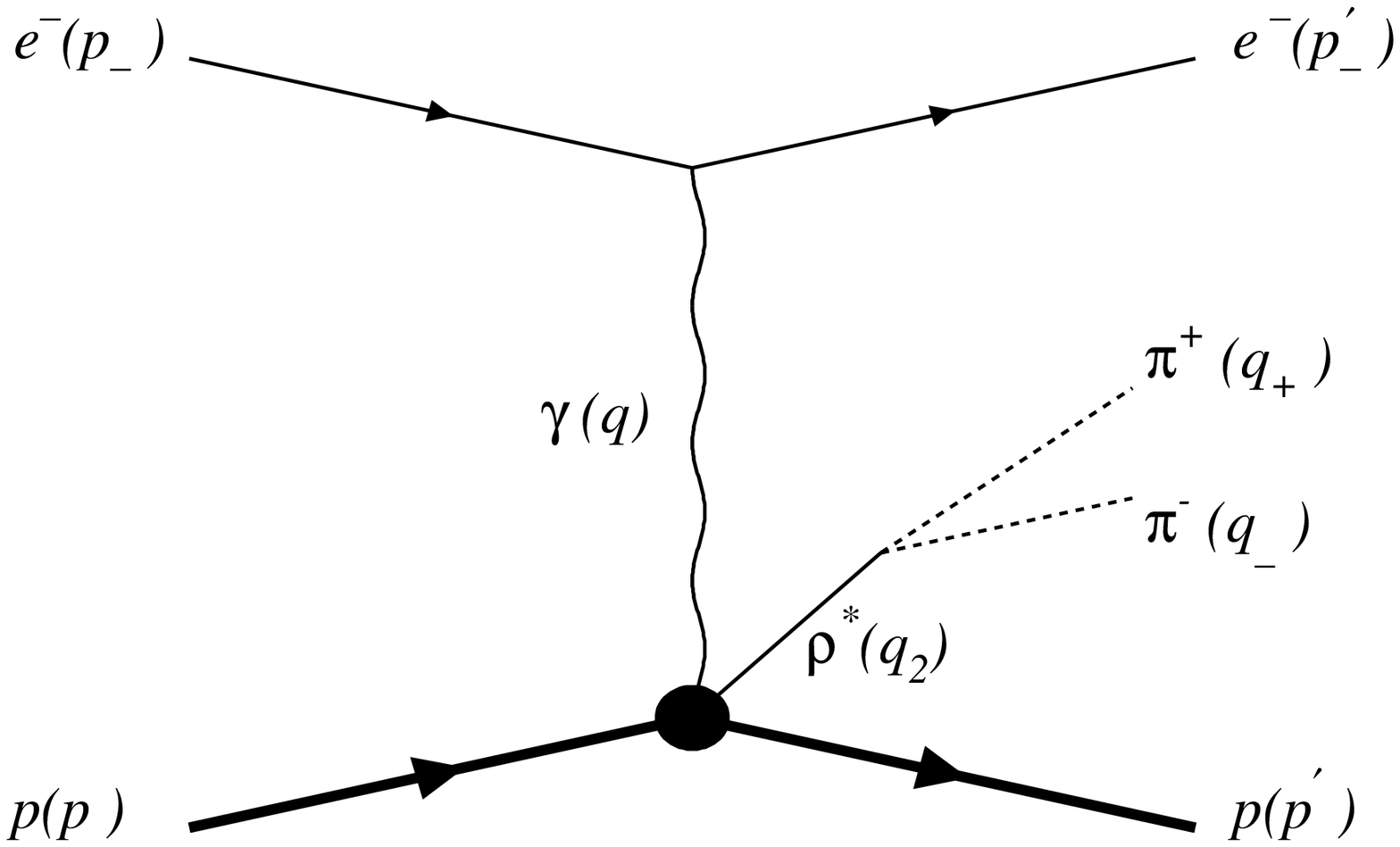}
\put(-50,10){\mbox{a)}}
}
\end{picture}
&
\begin{picture}(80,80)
\put(-5,0){
\epsfxsize=45mm
\epsfysize=45mm
\epsfbox{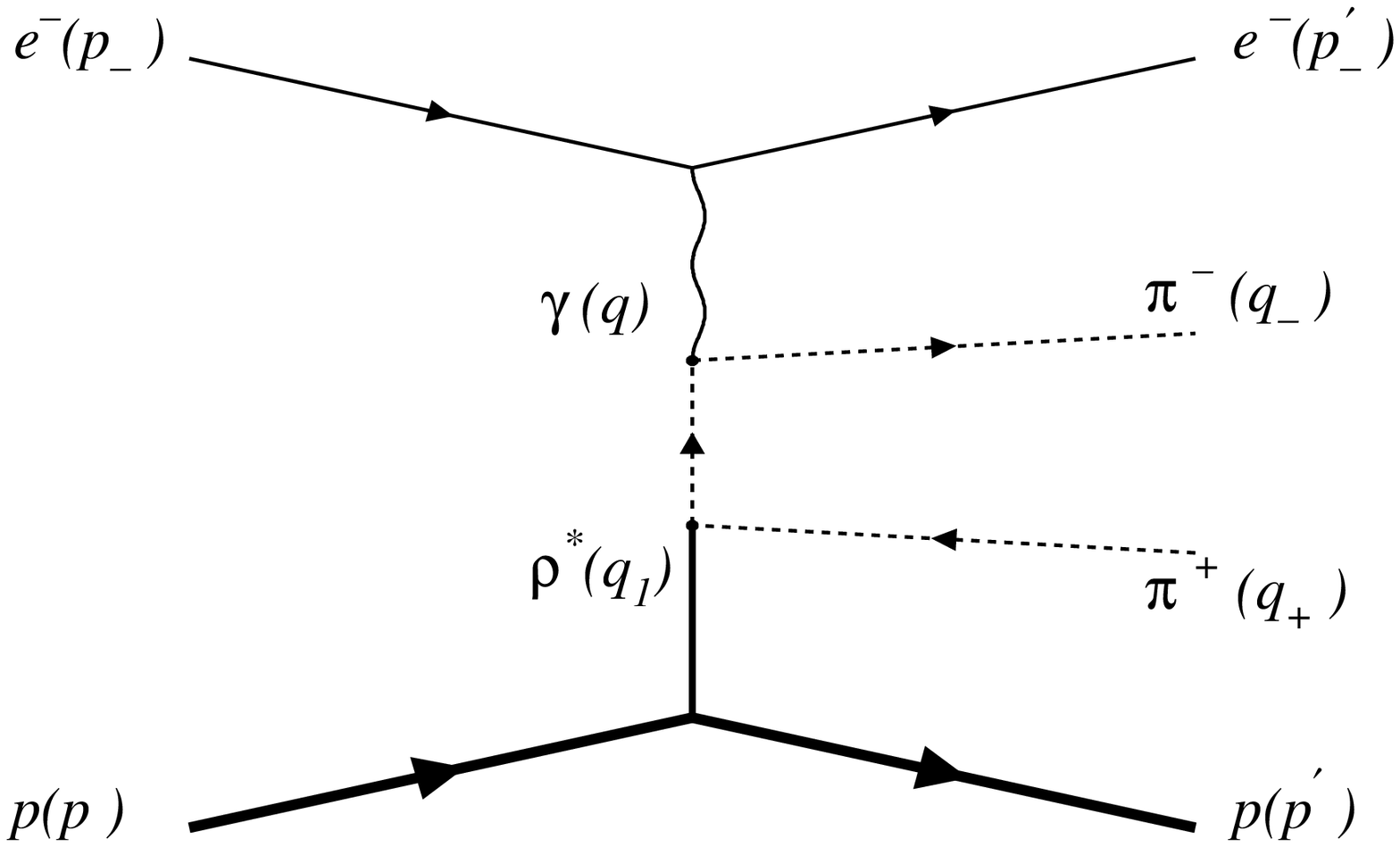}
\put(-50,10){\mbox{b)}}
}
\end{picture}
&
\begin{picture}(80,80)
\put(0,0){
\epsfxsize=45mm
\epsfysize=45mm
\epsfbox{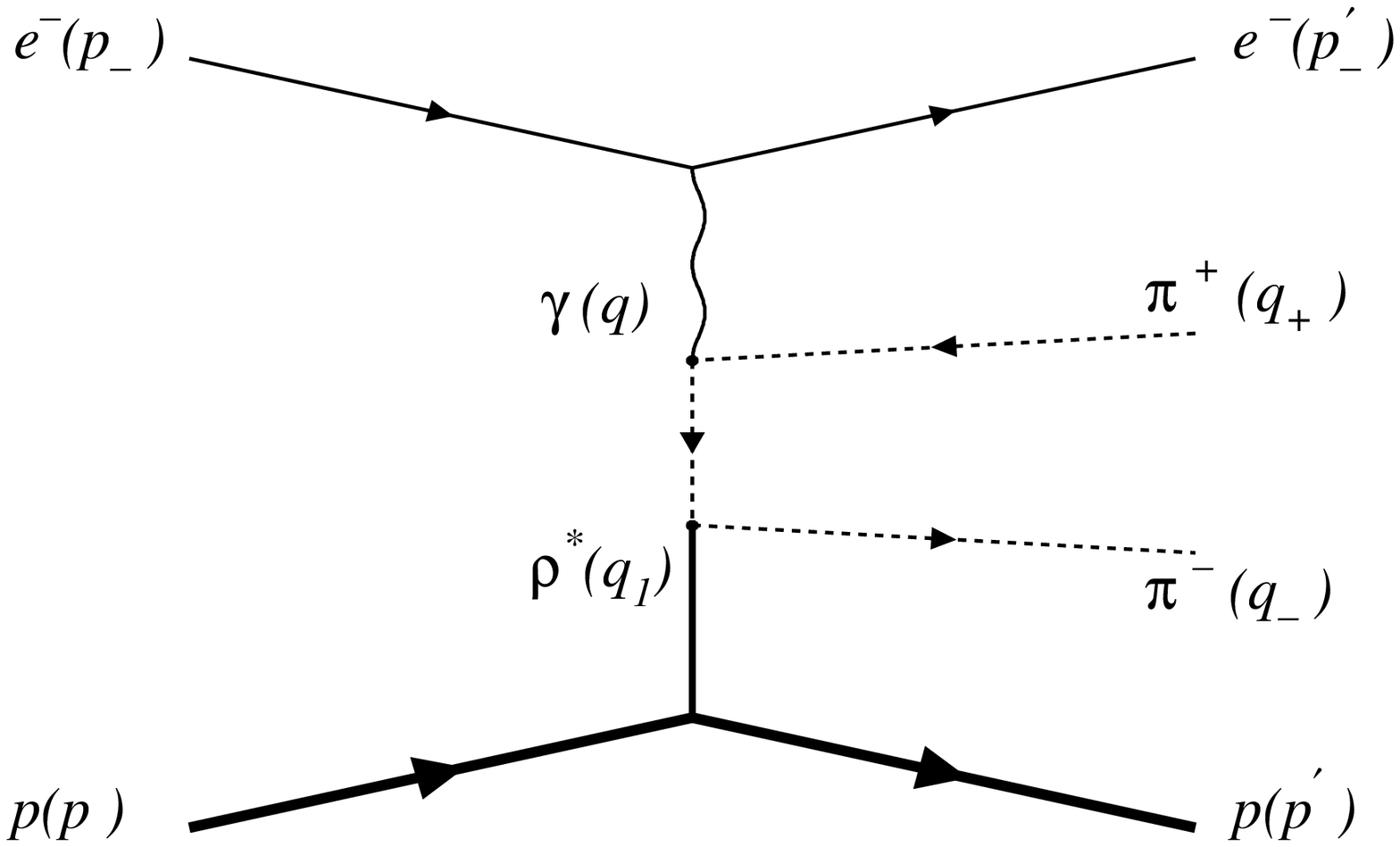}
\put(-50,10){\mbox{c)}}
}
\end{picture}
\\
\begin{picture}(80,80)
\put(-10,0){
\epsfxsize=45mm
\epsfysize=45mm
\epsfbox{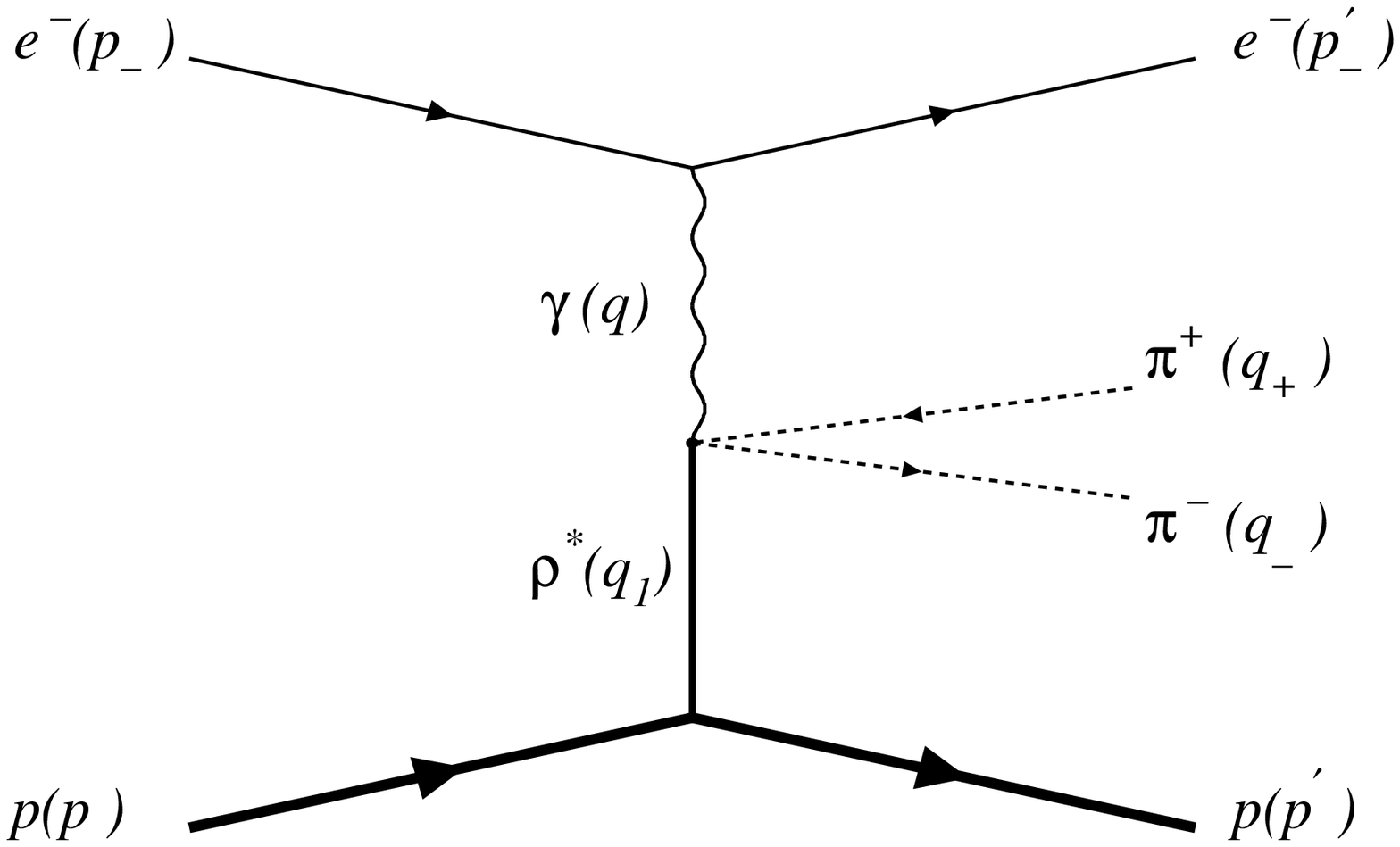}
\put(-50,10){\mbox{d)}}
}
\end{picture}
\end{tabular}
\caption{
Additional pion pair production in electron-proton scattering:
a) photon bftor meson and b)-d) bremsstrahlung mechanisms.
}
\label{fig:ppp}
\end{figure}
In this paper we propose to measure the odd part of DVCS
$\gamma p \to \gamma^* p$ and $\gamma p \to \rho^* p$, which is associated with
interference of two types of processes with
additional lepton
\ba
e(p_-)+p(p) \to e(p_-') + l^+(q_+) + l^-(q_-) + p(p'),
\ea
or pion
\ba
e(p_-)+p(p) \to e(p_-')+\pi^+(q_+)+\pi^-(q_-)+p(p')
\ea
pair production.
Namely we consider the interference of amplitudes of pair creation by so called
two-photon (Fig. \ref{fig:pll} a)) or photon $\rho $-meson (Fig. \ref{fig:ppp} a))
and bremsstrahlung (Fig. \ref{fig:pll} b)-e), Fig. \ref{fig:ppp} b)-d)) mechanisms.
The last one describe emission of
virtual photon (or vector meson) by proton with subsequent conversion to lepton or
pion pair.
To distinguish this mechanism from the other contributions to pair production processes
in electron-proton scattering we suggest to measure the charge-odd part of differential
cross section \cite{Ahmadov:2006if}
\ba
d\sigma_{odd}^{e^-p\to e^-\pi^+\pi^-p}&=&\sigma^\pi_0A^\pi(q_+,q_-)\frac{1}{M_\rho^4}d V_\pm \nn \\
d\sigma_{odd}^{e^-p\to e^-l^+l^-p}&=&\sigma^lA^l(q_+,q_-)\frac{1}{m^4}d V_\pm,
\ea
with $dV_\pm$ is the phase volume of the pair created and $A^{\pi,l}(q_+,q_-)$ is
some asymmetry function
\ba
A^{\pi,l}(q_+,q_-)=-A^{\pi,l}(q_-,q_+),
\ea
which incorporate besides QED as well some information about three current
correlation of the  proton block.

Really the process belongs to ones of higher orders on the fine structure constant
$\alpha$ but it is highly enhanced in the Weizsaecker-Williams (WW) approximation,
corresponding to the kinematics when the scattered electron moves to the direction
close to the it's initial one, which is partly compensate the extra power of
$\alpha$.

\section{Method of calculation and numerical results}
%
Below we will consider the case when proton state is not enhanced, remains to be proton.
The summed over the final and initial spin states
the interference of matrix elements for pion and lepton pair
production have a form:
\ba
\Delta \brm{M_\pi}^2 = 2M_\pi^a|M_\pi^b+M_\pi^c+M_\pi^d|^\dagger&=&
\frac{32(4\pi\alpha)^2(g_\pi g_N)^2}{(q^2)^2(q_1^2-M_\rho^2)(q_2^2-M_\rho^2)}
\times \nn\\&&
\times
T^e_{\mu\nu}T^P_{\eta\lambda\nu_1}T^\pi_{\mu_1\eta_1\lambda_1}G^{\mu\nu\lambda\eta\mu_1\nu_1\lambda_1\eta_1},
\ea
\ba
\Delta \brm{M_l}^2 = 2M_l^a|M_l^b+M_l^c|^\dagger&=&
\frac{128(4\pi\alpha)^4}{(q^2)^2q_1^2q_2^2}
T^e_{\mu\nu}T^P_{\eta\lambda\nu_1}T^l_{\mu_1\eta_1\lambda_1}G^{\mu\nu\lambda\eta\mu_1\nu_1\lambda_1\eta_1},
\ea
with
\ba
G^{\mu\nu\lambda\eta\mu_1\nu_1\lambda_1\eta_1}=g^{\mu\mu_1}g^{\nu\nu_1}g^{\lambda\lambda_1}g^{\eta\eta_1},
\ea
$g_\pi\approx g_N=10$ are coupling constants of $\rho$ meson with the charged pions and
nucleons, the transferred momentum $q=p_--p_-'$, $q_1=p-p'$, with $p_-'$, $p'$-the 4 momenta
of the scattered electron and proton, $q_2=q_++q_-$, $q_\pm$ is the momenta of pair created.
The contribution of the diagrams $M_l^d$ and $M_l^e$ are negligible since we consider
the kinematics of proton fragmentation which is relevant to DVCS measurement.

The cross sections then
\ba
\Delta d\sigma = \frac{1}{8s} \Delta \brm{M}^2 dV,
\ea
where phase volume $dV$ will be specified below.

\begin{figure}
\scalebox{1.}{\includegraphics{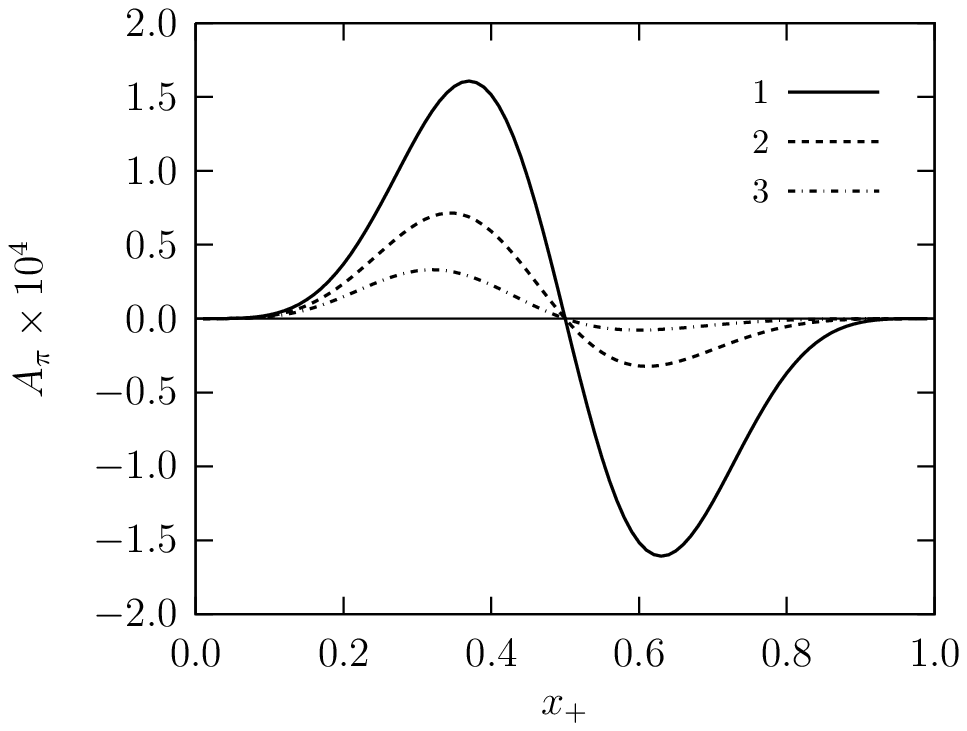}}
\caption{
The asymmetry (defined in (\ref{PionAsymmetry})) for pions pair production
as a function of $\pi^+$ energy fraction $x_+$.
Curve 1 corresponds to case then $\brm{\bf q_+}=\brm{\bf q_-}=2 M_\rho$,
Curve 2 corresponds to case then $\brm{\bf q_+}=2 M_\rho$, $\brm{\bf q_-}=2.5 M_\rho$,
Curve 3 corresponds to case then $\brm{\bf q_+}=2 M_\rho$, $\brm{\bf q_-}=3 M_\rho$.
}
\label{fig:Pion}
\end{figure}

\begin{figure}
\scalebox{1.1}{\includegraphics{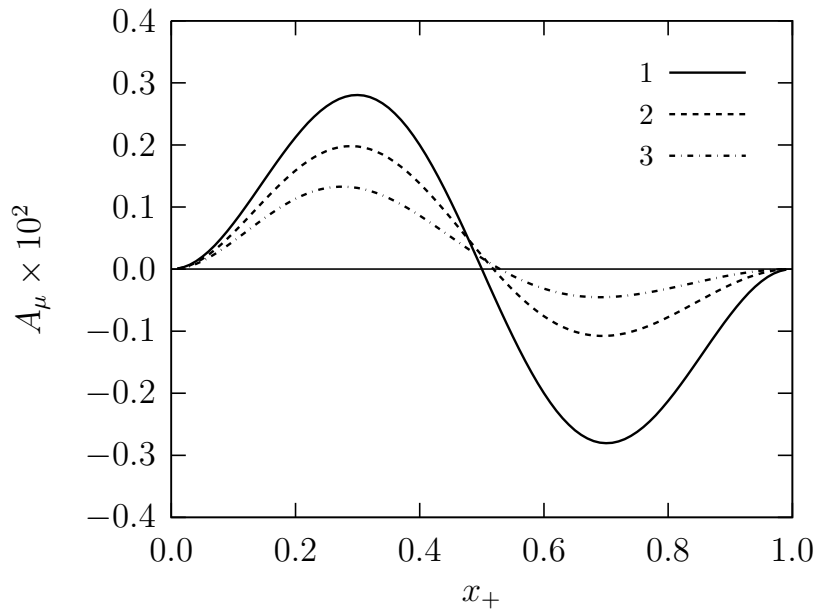}}
\caption{
The asymmetry (defined in (\ref{MuonAsymmetry})) for muons pair production
as a function of $\mu^+$ energy fraction $x_+$.
Curve 1 corresponds to case then $\brm{\bf q_+}=\brm{\bf q_-}=2 m_\mu$,
Curve 2 corresponds to case then $\brm{\bf q_+}=2 m_\mu$, $\brm{\bf q_-}=2.5 m_\mu$,
Curve 3 corresponds to case then $\brm{\bf q_+}=2 m_\mu$, $\brm{\bf q_-}=3 m_\mu$.
}
\label{fig:Muon}
\end{figure}

The electron tensor have a form:
\ba
T^e_{\mu\nu}=\frac{1}{4}{\rm Tr} \brs{\gamma_\mu\hat {p}_-\gamma_\nu\hat{p}_-'}.
\ea
Three index tensor with lepton pair is
\ba
T^l_{\mu_1\eta _1\lambda _1}&=&\frac{1}{4}{\rm Tr}
\Biggl[
(-\hat{q}_++m)\gamma_{\lambda _1} (\hat{q}_-+m)
\frac{\gamma_{\mu_1}(\hat {q}_--\hat{q}+m)\gamma_{\eta _1}}{D_-}+
\nn\\&&\;\;\;\;\;
+
(\hat{q}_-+m)\gamma_{\lambda _1} (-\hat{q}_++m)
\frac{\gamma_{\eta _1}(\hat{q}-\hat {q}_++m)\gamma_{\mu_1}}{D_+}\Biggr],  \\
D_\pm&=&(q-q_\pm)^2-m^2, \qquad q_\pm^2=m^2. \nn
\ea
Proton three indices tensor contains the virtual Compton tensor $O_{\lambda\sigma}$
and the vertex function of proton:
\ba
T^P_{\eta \lambda\nu _1}=\frac{1}{4}
{\rm Tr}\brs{(\hat{p}'+M)\Gamma_\eta (\hat{p}+M)O_{\lambda\nu _1}}. \nn
\ea
We will specify it for the case of point-like proton below.
The three index tensor with charged pions is
\ba
T^\pi_{\mu _1\eta	_1\lambda _1}&=&
\brs{\frac{\br{2q_--q}_{\mu_1}\br{q_1-2q_+}_{\eta _1}}{D_-}+
\frac{\br{2q_--q_1}_{\eta _1}\br{q-2q_+}_{\mu_1}}{D_+}-2g_{\mu _1\eta _1}}
\times\nn\\&&
\times
\br{q_--q_+}_{\lambda_1}.
\ea
In the peripheral kinematics
\ba
s=(p_-+p)^2 \gg q_2^2 \gg |q^2|
\ea
(which provide the maximal contribution to the cross section)
the substitution
\ba
G^{\mu\nu\lambda\eta\mu_1\nu_1\lambda_1\eta_1}=(2/s)^4p^\mu p^\nu p^{\lambda_1} p^{\eta_1} p_-^{\mu_1}p_-^{\nu_1}
p_-^{\lambda}p_-^{\eta},
\ea
is valid.
Throughout the paper we will use Sudakov's \cite{Sudakov:1954sw,Baier:1980kx}
(infinite momentum frame) parametrization
of momenta of the problem:
\ba
q_1&=&\alpha_1 \bar{p}+\beta_1\bar{p}_-+q_{1\bot},\nn\\
q&=&\alpha \bar{p}+\beta\bar{p}_-+q_{\bot}, \nn \\
q_\pm&=&\alpha_\pm\bar{p}+\beta_\pm\bar{p}_-+q_{\pm\bot}; \nn \\
p'&=&\alpha'\bar{p}+\beta'\bar{p}_-+p'_\bot,
\ea
where we define
\ba
q_\bot p_-=q_\bot p=0, \qquad q_\bot^2=-{\bf q}^2,
\ea
and $\bar{p}_-,\bar{p}$ are the light-like 4-vectors
\ba
\bar{p}_- = p_- - p \frac{m_e^2}{s},
\qquad
\bar{p} = p - p_- \frac{M^2}{s},
\qquad
2\bar{p}_-\bar{p}=s.
\ea
We use below the on mass shell conditions
\ba
s\alpha_\pm=\frac{1}{\beta_\pm}[{\bf q}_\pm^2+m^2],
\qquad s\beta'=\frac{1}{\alpha'}[{\bf p'}_1^2+M^2]
\ea
and conservation law
\ba
{\bf q}_1={\bf q}_++{\bf q}_-; \qquad
\alpha_1=\alpha_++\alpha_-; \qquad
\beta=\beta_++\beta_-,
\ea
(we use here the WW condition $|{\bf q}| \ll {\bf q}_\pm\sim |{\bf q}_1|$ and the
kinematic features of peripheric interaction.
The relevant kinematic invariants are:
\ba
q_1^2&=&-{\bf q_1^2}=-({\bf q}_++{\bf q}_-)^2; \nn\\
q_2^2&=&(q_++q_-)^2=
\frac{1}{x_+x_-}[m^2+(x_-{\bf q}_+-x_+{\bf q}_-)^2],\nn \\
x_\pm&=&\frac{\beta_\pm}{\beta}=\frac{pq_\pm}{pq}, \qquad
0 < x_\pm < 1, \qquad x_++x_-=1.
\ea
Besides we use
\ba
d^4q=\frac{s}{2}d\alpha d\beta d^2{\bf q}.
\ea
Factor $G$ provides the light-cone projection of tensor product:
\ba
T^e_{\mu\nu}T^P_{\eta\lambda\nu_1}T^{\pi,l}_{\mu_1\eta_1\lambda_1}G^{\mu\nu\lambda\eta\mu_1\nu_1\lambda_1\eta_1}=
\br{\frac{2}{s}}^4s^2N_e sN^{\pi,l}s^3N^p\sim s^2,
\ea
with finite in the limit $s\to \infty$ quantities $N_e,N^p,N^\pi,N^l$.
We had convinced that the
matrix element squared is proportional $s^2$. Besides we will see that it is
proportional to ${\bf q}^2$. This fact is the consequence of gauge invariance.
The explicit calculations give:
\ba
N_e=\frac{1}{4s^2}{\rm Tr}\brs{\hat{p}\hat{p}_-\hat{p}\hat{p}'_-}=\frac{1}{2}.
\ea
The quantity $N^p$ we will calculate in point like proton approximation:
\ba
N^p=\frac{1}{4s^3}{\rm Tr}\Biggl [
\br{\hat{p}+M}
\Biggl (
    \frac{\hat{p}_-\br{\hat{p}+\hat{q}+M}\hat{p}_-}{D}+
\nn \\
    \frac{\hat{p}_-\br{\hat{p}'-\hat{q}+M}\hat{p}_-}{D'} \Biggl )
    \br{\hat{p}'+M}\hat{p}_- \Biggr ],
\ea
with $D=(p+q)^2-M^2\approx s\beta$ and $D'=(p'-q)^2-M^2\approx -s\alpha'\beta+2{\bf p}'{\bf q}$.
The result is:
\ba
N^p=\frac{\alpha'}{2}\brs{\frac{1}{D}+\frac{\alpha'}{D'}}=\frac{{\bf q}_1{\bf q}}{(s\beta)^2}.
\ea
The expression for $N^l$ is
\ba
N^l=\frac{1}{4s}
{\rm Tr}\Biggl[
\br{-\hat{q}_++m}\hat{\bar{p}}\br{\hat{q}_-+m}
\frac{\hat{p}_-\br{\hat{q}_--\hat{q}+m}\hat{\bar{p}}}{D_-}+
\nn\\+
\br{\hat{q}_-+m}\hat{\bar{p}}\br{-\hat{q}_++m}
\frac{\hat{\bar{p}}\br{\hat{q}-\hat{q}_++m}\hat{p}_-}{D_+}\Biggr].
\ea
Using the current conservation conditions
\ba
T^{l,\pi}_{\mu\nu\lambda}q^\mu=T^{l,\pi}_{\mu\nu\lambda}q_1^\nu=0
\ea
and
\ba
q&\approx&\beta\bar{p}_-+q_\bot,\nn\\
q_1& \approx& \alpha_1 \bar p + q_{1\bot}, \nn
\ea
we can replace $p_- \to-q_\bot/\beta$ and $\bar{p} \to -q_{1\bot}/\alpha_1$.
Simple calculation lead to (we use $D_\pm=-d_\pm/x_\pm, d_\pm={\bf q}_\pm^2+m^2$):
\ba
N^l=\frac{s\beta}{s_1}
\brs{x_+\frac{{\bf q}_+{\bf q}}{d_+}-x_-\frac{{\bf q}_-{\bf q}}{d_-}}
\brs{x_-\br{{\bf q}_1{\bf q}_+}+x_+\br{{\bf q}_1{\bf q}_-}},
\ea
For the case of pion pair production we have
\ba
N^\pi=\frac{s\beta\br{x_--x_+}}{s_1}
\brs{\br{d_+-d_-}\br{x_-\frac{{\bf q}_-{\bf q}}{d_-}-
x_+\frac{{\bf q}_+{\bf q}}{d_+}}+{\bf q}{\bf q}_1}.
\ea
The factor ${\bf q}_i{\bf q}_j$ must be
extracted, which after angular averaging turns to $1/2 \delta_{ij}{\bf q}^2$
and the rest part must be evaluated at ${\bf q}=0$.
Next step consists in transformation of phase volume. Introducing as a variables
the momentum transferred
\ba
d^4q_1\delta^4(p-q_1-p') d^4q\delta^4(p_--p_-'-q), \nn
\ea
and using the Sudakov parametrization we transform the phase volume
\ba
dV=(2\pi)^{4-12}\delta^4(p_-+p-p_-'-p'-q_+-q_-)\frac{d^3p_-'}{2E'_-}\frac{d^3p'}{2E'}
\frac{d^3q_-}{2E_-}\frac{d^3q_+}{2E_+},
\ea
as
\ba
dV&=&\frac{(2\pi)^{-8}\pi^3}{8s}\frac{d^2{\bf q}}{\pi}\frac{d\beta}{\beta}dV_\pm, \nn \\
dV_\pm&=&\frac{d^2{\bf q_+}}{\pi}\frac{d^2{\bf q_-}}{\pi}
\frac{dx_-}{x_+x_-}.
\ea
Employing the form of the photon momentum squared by electron $q^2$:
\ba
q^2\approx-\brs{{\bf q}^2+\beta^2m_e^2},
\ea
\ba
\int\limits_0^{Q^2}\frac{{\bf q}^2 d{\bf q}^2}{\br{{\bf q}^2+m_e^2\beta^2}^2}=
\ln\br{\frac{Q^2 s^2}{m_e^2s_1^2}}-2\ln\br{s\beta/s_1}-1,
\ea
where $Q^2 \sim M_p^2$.
Further integration on $s\beta>s_1={\bf q}_+^2/x_++{\bf q}_-^2/x_-$ leads to
\ba
\frac{1}{s_1}R = \int\limits_{s_1}^\infty\brs{\ln\br{\frac{Q^2 s^2}{m_e^2s_1^2}}-2\ln{s\beta/}{s_1}-1}
\frac{d(s\beta)}{(s\beta)^2}=
\frac{1}{s_1}\brs{\ln\frac{Q^2s^2}{m_e^2s_1^2}-3}.
\ea
The final results for odd part of cross sections are:
\ba
\Delta d\sigma^{ep\to e\pi_+\pi_-p}&=&\sigma_0^\pi \frac{A_\pi}{x_+x_-} \frac{x_+x_-dV_\pm}{M_\rho^4}, \nn \\
\sigma_0^\pi&=&\frac{\alpha^2(g_\rho g_N)^2}{8\pi^3M_\rho^2} = 1.4 \mu b; \nn \\
A_\pi &=&(x_--x_+)\frac{M_\rho^6(q_2^2-M_\rho^2)R}{({\bf q}_1^2+M_\rho^2)
[(q_2^2-M_\rho^2)^2+M_\rho^2\Gamma_\rho^2]} \times \nn\\
&\times&
\frac{1}{s_1^2d_+d_-}[(d_+-d_-)(x_-d_+{\bf q}_-{\bf q}_1-x_+d_-{\bf q}_+{\bf q}_1)
+
\nn\\
&+&
d_+d_-{\bf q}_1^2],
\label{PionAsymmetry}
\ea
with
\ba
s_1 = \frac{d_+}{x_+} + \frac{d_-}{x_-},
\qquad
q_2^2 = \frac{1}{x_+ x_-}
\br{
    \br{x_- {\bf q_+} - x_+ {\bf q_-}}^2 + M_\pi^2
},
\ea
and
\ba
\Delta d\sigma^{ep\to e\l_+l_-p}&=&\sigma_0^l \frac{A_l}{x_+x_-} \frac{x_+x_-dV_\pm}{m^4},
\qquad
\sigma_0^l=\frac{8\alpha^4}{\pi m^2} = 0.2 \nb~~ \mbox{for}~ m=M_\mu, \nn \\
A_l&=&-\frac{m^6R}{s_1^2d_+d_-{\bf q}_1^2q_2^2}
\brs{x_+d_-\br{{\bf q}_1{\bf q}_+}-x_-d_+\br{{\bf q}_1{\bf q}_-}}
\times \nn\\&\times &
\brs{x_-\br{{\bf q}_1{\bf q}_+}+x_+\br{{\bf q}_1{\bf q}_-}}
\label{MuonAsymmetry}.
\ea
The functions $A^\pi/\br{x_+x_-}$, $A^l/\br{x_+x_-}$ are drawn in
Fig.~\ref{fig:Pion}, \ref{fig:Muon} for specific values of
${\bf q}_\pm$ as a function of $x_+$.

\section{Conclusion}

We investigate the light-cone projection of DVCS rank 3 tensor for
specific processes of pair production.
The value of charge-odd contributions to the cross sections
(see (\ref{MuonAsymmetry}), (\ref{PionAsymmetry})) are sufficiently large to be measured.
It can be measured in experiment with inclusive pion and lepton pairs detection.

\section{Acknowledgements}
We thank Dr. A.~Nagaytsev and Prof. V.G.~Krivokhizhin for interest to this problem.


\end{document}